\documentclass{appolb}
\usepackage{graphicx}
\usepackage{amsmath}

\begin{document}
\title{A QFT scalar toy model analogous to positronium and pion decays%
\thanks{Based on the presentation of M.P. at the Symposium on `New Trends in Nuclear and Medical Physics', 18-20 October 2023, Cracow, Poland}%
}
\author{M. Piotrowska$^{1}$, F. Giacosa$^{1,2}$
\address{$^1$ Institute of Physics, Jan Kochanowski University,\\ \textit{ul. Uniwersytecka 7, 25-406 Kielce, Poland.}\\
$^{2}$ Institute for Theoretical Physics, J. W. Goethe University, \\ \textit{Max-von-Laue-Str. 1, 60438 Frankfurt, Germany.}}
\\[3mm]
}
\maketitle
\begin{abstract}
In the framework of a scalar QFT, we evaluate the decay of an initial massive state into two massless particles through a triangle-shaped diagram in which virtual fields propagate.
Under certain conditions, the decaying state can be seen as a bound state, thus it is analogous to the neutral pion (quark-antiquark pair) and to the positronium (electron-positron pair), which decay into two photons. While the pion is a relativistic composite object, the positronium is a non-relativistic compound close to threshold. We exam similarities and differences between these two types of bound states.

\end{abstract}
  
\section{Introduction}
The positronium, being the bound state of one electron and one positron, is one of the simplest systems described by Quantum Electrodynamics (QED) \cite{Bass:2023dmv,Bass:2019ibo}. Besides its relevance in fundamental physics, its applications  in medical physics is recognized \cite{moskal2019nrp}.
The lightest positronium, named para-positronium (or positronium tout-court below) is the $1 ^1S_0$ ground-state with anti-parallel spins.  This state is  not stable, but it decays into two photons with a lifetime of about 0.12 ns \cite{Czarnecki:1999ci,Kniehl:2000dh,Czarnecki:1999mt}. 
Interestingly, in the realm of Quantum Chromodynamics (QCD), there is an object analogous to the positronium:  the neutral pion, which is a quark-antiquark $1 ^1S_0$ bound state decaying into two photons in about $8.43 \times 10^{-17}$ s \cite{pdg}. 

Yet, besides these similarities, the positronium and the pion differ significantly from each other for important reasons. The former is a loosely non-relativistic bound state very close to the threshold, the latter is a deeply bound relativistic object that plays the role of a quasi-Goldstone boson of QCD. 

Nevertheless, we argue here that a similar process is at the basis of their decays, namely a triangle-shaped loop of electrons and quarks, respectively. In order to discuss this point of view, we construct a Quantum Field Theoretical (QFT) scalar toy model that, under some specific conditions, shares some similarities with the two-photon decay of both the positronium and the neutral pion. 
The aim of this study is twofold: first, an interesting aspect is what can be learned about the QFT approach from the comparison with the well-known and precise positronium results; second, we wonder if the QFT approach can deliver some novel insights on the non-relativistic positronium case. As a cautionary remark, we stress that we do not aim to compete with the precision of QED, but rather to establish a connection between these systems. 
Moreover, here we do not yet work with the `real case' since we stick to scalar fields only, which serve to highlight the main aspects of the problem.

In particular, we shall introduce our model by fixing the Lagrangian and discussing under which conditions it is suitable for the description of bound state, such as  positronium or pion (via the so-called Weinberg compositeness condition \cite{Weinberg,Hayashi}). Moreover, the evaluation of the triangle-shaped diagram with massive scalars circulating in it (analogous to fermions) leading to decays into massless scalar fields (analogous to photons) allows for a simple (and in a certain sense didactic) introduction to the main features of the two-photon decays of the positronium and the pion.

\section{Scalar model: loop-driven decay}
We introduce the interaction Lagrangian as
\begin{equation}
\mathcal{L}_{int}=\frac{g_{S}}{2}S\varphi ^{2}+\frac{g_{A}}{2}A\varphi ^{2} ,
\end{equation}
with the scalar field $S(x)$ (analogous to the positronium or the pion) with  mass $M_S$, the (pseudo)scalar field $\varphi(x)$ (analogous to the electron or the quark) with mass $m_{\varphi}$, and the massless scalar field $A(x)$ (analogous to the photon). 
Moreover, $g_S$ and $g_A$ stand for the coupling constants (the latter being analogous to the electric charge).
In the following, we set $m_{\varphi} = 1$ (alias, $m_{\varphi}$ serves as a unit of energy for the model).

\begin{figure}[htb]
\centerline{%
\includegraphics[width=0.5\linewidth]{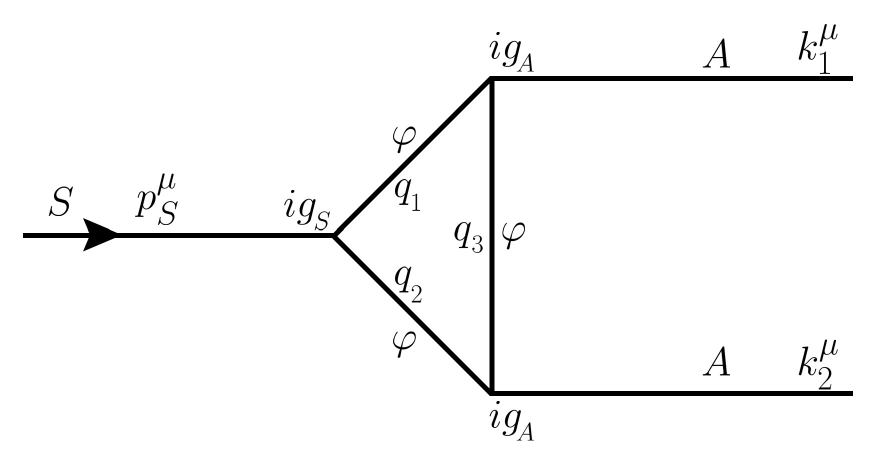}}
\caption{Schematic illustration of the process $S\rightarrow AA$.}
\label{fig:decay}
\end{figure}

The scalar field $S$ does not couple directly to $AA$, but this decay is possible via a triangle diagram involving $\varphi$-particles, see Fig. \ref{fig:decay}. Our goal is to calculate $S \rightarrow AA$. Note, we assume that $m_S<2m_{\varphi}$, thus the decay $S \rightarrow \varphi \varphi$ is not kinematically allowed. This condition is surely met when interpreting $S$ as a $\varphi \varphi$ bound state. 

Next, we briefly present the kinematics. The external momenta of the diagram of Fig. \ref{fig:decay} are as follow: $p^{\mu}_S=(M_S, \vec{0})$, $k_1^{\mu}=(E_A, 0, 0, k_z)$~and $k_2^{\mu}=(E_A, 0, 0, -k_z)$, with $E_A=k_z, k_z>0$ and $E_A=M_S/2$. For what concerns the internal momenta one has: $q_{1}=\frac{p}{2}+q$, $q_{2}=\frac{p}{2}-q$ and $q_{3}=\frac{p}{2}+q-k_{1}$. 

A central object is the amplitude $I$ associated with the triangle diagram of Fig. \ref{fig:decay}:
\begin{equation}
 I=\int \frac{d^{4}q}{(2\pi )^{4}}\frac{ \phi(\vec{q})}{(q_1^2-m_{\varphi}^2+ i\varepsilon)(q_2^2-m_{\varphi}^2+ i\varepsilon)(q_3^2-m_{\varphi}^2+ i\varepsilon)}
 \text{ ,}
 \label{amplitude}
\end{equation}
where  $\phi(\vec{q})=e^{\vec{q}^2/\Lambda^2}$(with the choice $\Lambda=1$) is a vertex function (chosen to be real for simplicity) associated with the $S \varphi \varphi$ nonlocal interaction \cite{Faessler:2003yf,Giacosa:2007bs,Giacosa:2004ug} (for a discussion of covariance, see \cite{Soltysiak:2016xqz}).
In the following, we present two different ways of evaluating the above integral. The first one makes use of the Wick rotation, while the second is based on the residue theorem. The latter allows also to identify the dominating contribution in the non-relativistic limit. 

Let us rewrite the triangle amplitude of Eq. (\ref{amplitude}) as
$I=\int \frac{d^{4}q}{(2\pi )^{4}}\frac{\phi(\vec{q})}{D_1D_2D_3}$ with: 
\begin{eqnarray}
\label{d12}
&D_{1,2}&=\left(p/2 \pm q\right)^2-m_{\varphi}^2+i \varepsilon=\left(M_S/2 \pm q^0\right)^2-\vec{q}^2-m_{\varphi}^2+i \varepsilon, \nonumber \\ 
&D_3&=\left(M_S/2+q^0-k_1^0\right)^2-(\vec{q}-\vec{k_1})^2-m_{\varphi}^2+i \varepsilon \text{ .} 
\end{eqnarray}
The Wick rotation method amounts to the substitution $q^0=iw$, thus the integral is performed along the vertical axis on the complex plane, which can be evaluated numerically. 
Within the  residue approach, one performs the integral over $q^0$ analytically: 
$\int dq^4 = \int d^3q \int dq^0 \stackrel{\mathrm{Residue \  theorem}}{=} \int d^3q$,
and the remaining integral numerically. 
The poles corresponding to $D_{1,2,3}=0$, see Eq. (\ref{d12}), are listed in Table \ref{poles}. One can then write $ D_k=(q^0-L_k)(q^0-R_k)$ with $k=1,2,3$.
 \begin{table}[h!]
\renewcommand{\arraystretch}{1.4}
\centering
\begin{tabular}{|c|c|}
\hline
Poles of $D_1$ & Poles of $D_2$  \\ \hline
$L_1=-\frac{M_S}{2}-\sqrt{\rho^2+q_z^2+m_{\varphi}^2}+i\delta$& $L_2=\frac{M_S}{2}-\sqrt{\rho^2+q_z^2+m_{\varphi}^2}+i\delta$\\
$R_1=-\frac{M_S}{2}+\sqrt{\rho^2+q_z^2+m_{\varphi}^2}-i\delta$&$R_2=\frac{M_S}{2}+\sqrt{\rho^2+q_z^2+m_{\varphi}^2}-i\delta$\\
\hline
\multicolumn{2}{|c|}{Poles of $D_3$}\\
\hline
\multicolumn{2}{|c|}{$L_3=-\sqrt{\rho^2+(q_z-k_z)^2+m_{\varphi}^2}+i\delta$}\\
\multicolumn{2}{|c|}{$R_3=\sqrt{\rho^2+(q_z-k_z)^2+m_{\varphi}^2}-i\delta$}\\
\hline
\end{tabular}
\caption{ \label{poles} Analytical formulas for the poles. Note, $q_x^2+q_y^2=\rho^2$.}
\end{table}

Finally, the total decay width for the process $S\rightarrow AA$ takes the following form: 
\begin{equation}
\Gamma_{S \rightarrow AA}=\frac{1}{2}\frac{|\vec{k}|}{8 \pi M_S^2}\left| 2ig_{S\varphi \varphi}(2ig_{A \varphi \varphi})^2 I\right|^2=\frac{4|\vec{k}|}{ \pi M_S^2}g^2_{S \psi \psi}g^4_{A \varphi \varphi} \left| I \right| ^2 
    \text{ .}
\end{equation}

\begin{figure}[htb]
\centerline{%
\includegraphics[width=0.6\linewidth]{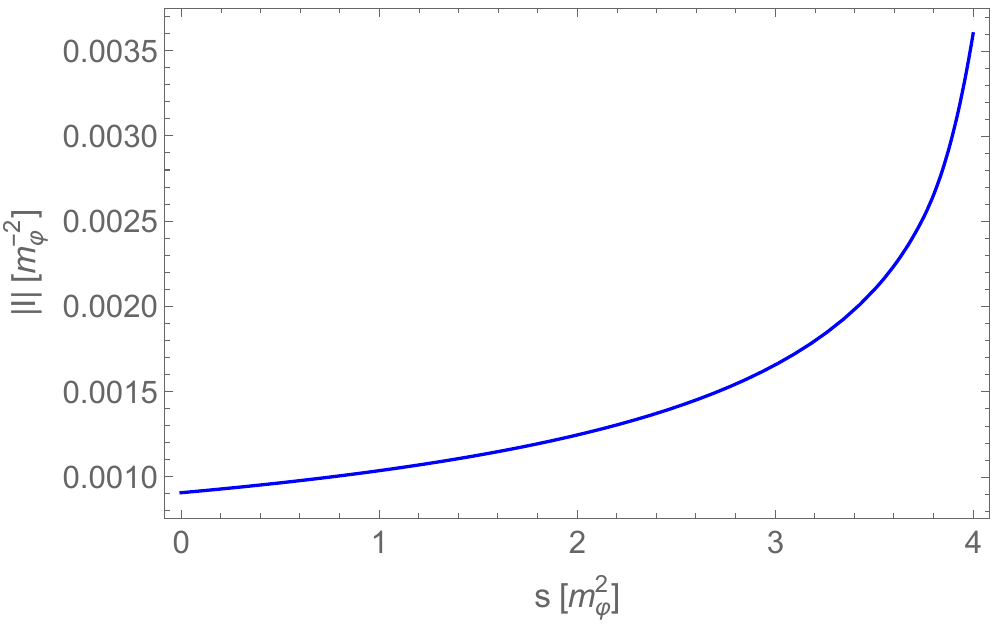}}
\caption{Behavior of the triangle amplitude $I$ of Eq. (\ref{amplitude}) as function of invariant variable $s=M_S^2$.}
\label{fig:WRmodel1}
\end{figure}

The results for the amplitude $I$ of Eq. (\ref{amplitude}) are shown in Fig. \ref{fig:WRmodel1}. 
For increasing values of $s$, the amplitude $|I|$ also increases and has a cusp at the threshold.
\begin{table}[h!]
\renewcommand{\arraystretch}{1.0}
\centering
\begin{tabular}{|c|c|c|}
\cline {2 -3}
\multicolumn{1}{ c |}{} & \multicolumn{1}{ c|}{RELATIVISTIC ($M_S=1$)}& \multicolumn{1}{ c|}{NON-RELATIVISTIC ($M_S=1.99$)} \\
\multicolumn{1}{ c |}{} & \multicolumn{1}{ c|}{(pion like)}& \multicolumn{1}{ c|}{(positronium like)} \\
\cline {2 -3}
\multicolumn{1}{c |}{} & \multicolumn{2}{c |}{Residue theorem}\\
\hline
$|I_1|$ & $0.0039131$ & $0.00410328$ \\ \hline
$|I_2|$ & $0.00176375$ & $0.000352004$ \\ \hline
$|I_3|$ & $0.00464127$ & $0.00104738$ \\ \hline
$|I|$ & $0.00103557$ & $0.0034079$ \\ \hline
\multicolumn{1}{c |}{} & \multicolumn{2}{c |}{Wick rotation method}\\
\hline
$|I|$ & $0.00103557$ & $0.0034079$ \\ \hline
\end{tabular}
\caption{\label{pionpos} Numerical results for the amplitude $I$ of Eq. (\ref{amplitude}) for chosen mass values for the decaying particle $S$. $I$ is in units of $m_{\varphi}^{-2}$.}
\end{table}
 Moreover, as an illustrative example, in Table \ref{pionpos} we present numerical results  for two mass values of the $S$ state. They were chosen as follow: $M_S=1$, deeply bound (thus `pion' like), and $M_S=1.99$, loosely bound  (hence `positronium' like). 
While in the former case, the contributions from all three poles are significant, for the latter the first `non-relativistic' pole dominates.

\begin{figure}[htb]
\centerline{%
\includegraphics[width=0.52\linewidth]{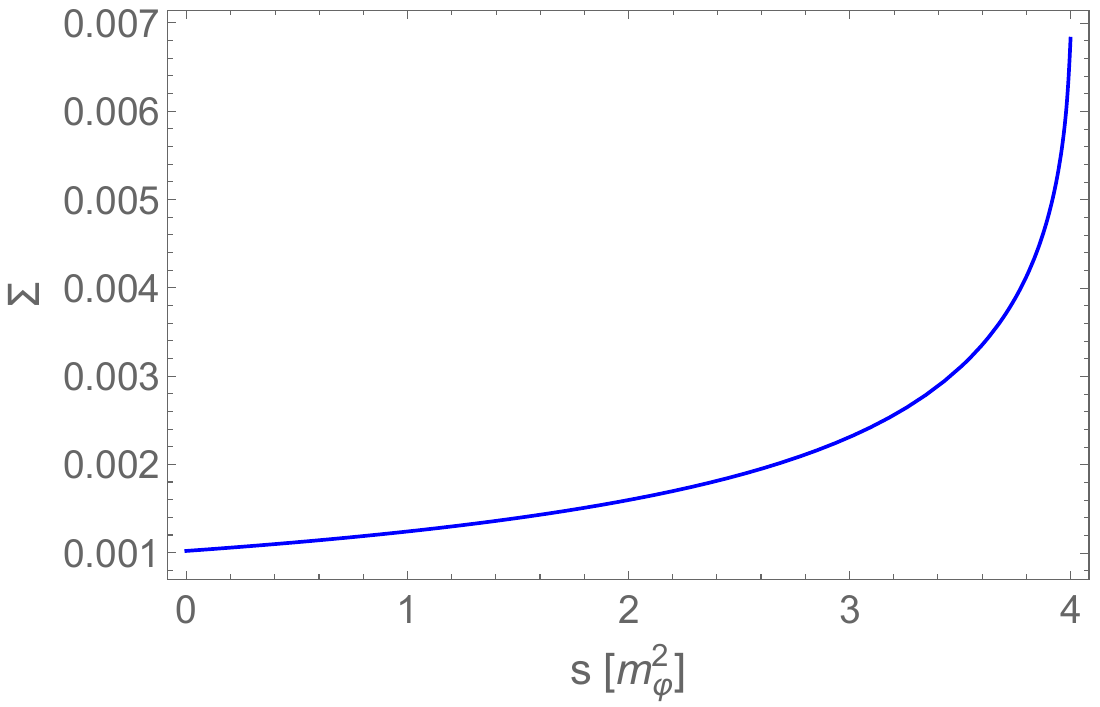}\includegraphics[width=0.5\linewidth]{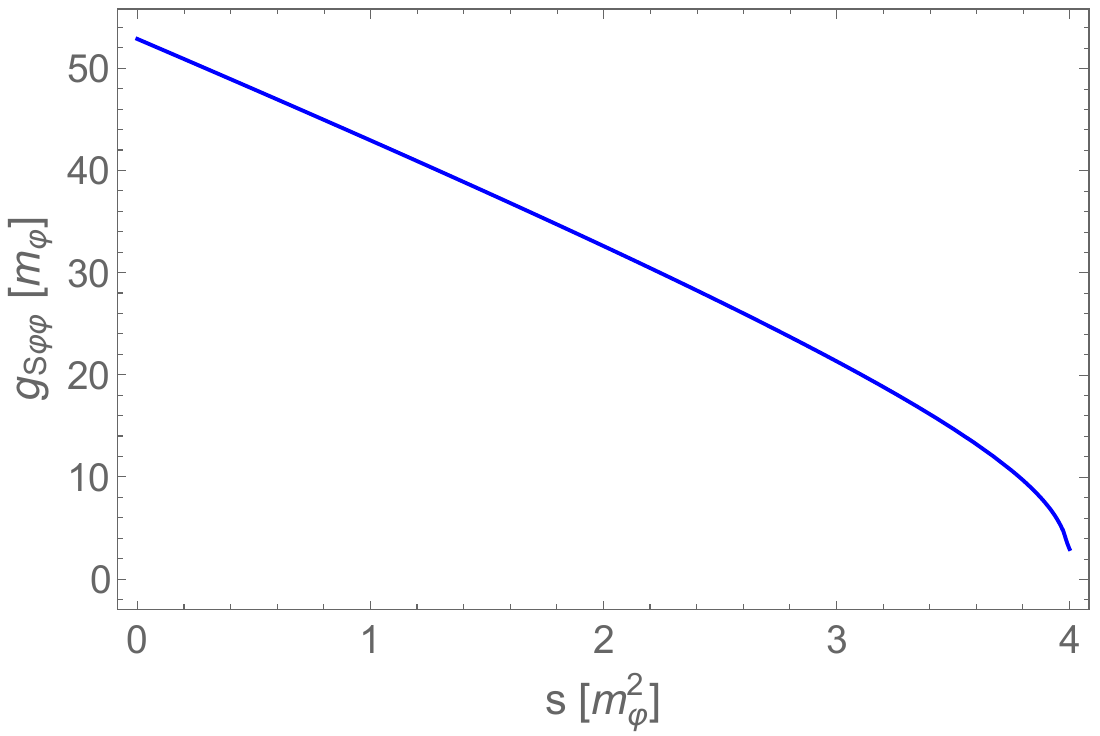}}
\caption{Left: loop function $\Sigma(s=M_S^2)$;  right: coupling constant $g_{S\varphi \varphi}$ as function of $s=M_S^2$. }
\label{fig:WRmodel3}
\end{figure}

When the state $S$ represents a $\varphi \varphi$ bound state, the coupling constant $g_{S\varphi \varphi}$ is not a free parameter any longer, but can be calculated via the Weinberg's compositeness condition as $ g_{S\varphi \varphi}= \left(2\Sigma'(s=M_S^2)\right)^{-1/2}$ \cite{Weinberg,Hayashi} (see also \cite{Faessler:2003yf,Giacosa:2007bs} for dealing explicitly with the pion case), where the loop function $\Sigma(s) = -i\int \frac{d^{4}q}{(2\pi )^{4}}\frac{\phi(\vec{q})^2}{D_1D_2} $ describes the process $S \rightarrow \varphi \varphi \rightarrow S$. The vertex function $\phi(\vec{q})$ is proportional to the wave function of the bound state in momentum space.
The loop function $\Sigma (s)$ as well as the coupling constant $g_{S\varphi \varphi}(s)$ are presented in Fig. \ref{fig:WRmodel3}.
While $\Sigma(s)$ increases when $s$ increases, the coupling constant $g_{S\varphi \varphi}$ decreases and vanishes at threshold. As a consequence, the coupling in the deep bound case ($M_S=1$) turns out to be one order of magnitude larger than the loosely bound case ($M_S=1.99$).

\begin{figure}[htb]
\centerline{%
\includegraphics[width=0.5\linewidth]{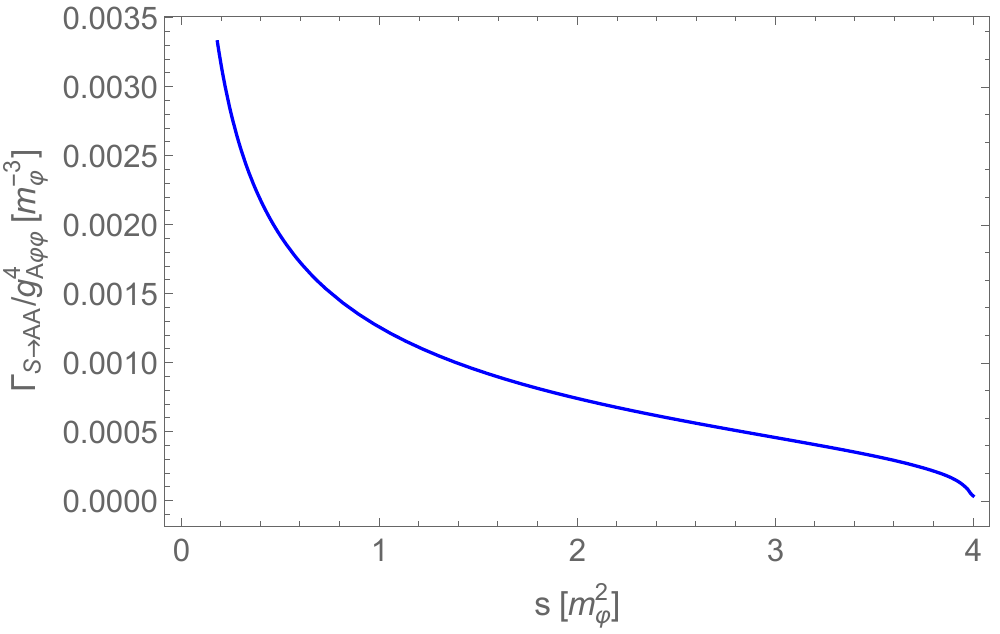}}
\caption{Decay width $\Gamma_{S \rightarrow AA}$ (normalized to $g_{A\varphi \varphi}^4$) as function of $s=M_S^2$.}
\label{fig:gamma}
\end{figure}

As a final result, in Fig. \ref{fig:gamma} we present the decay width $\Gamma_{S \rightarrow AA}$ (normalized to $g_{A\varphi \varphi}^4$) as function of the running squared mass $s=M_S^2$. One sees that it decreases for increasing $M_S$ and tends to 0 close to threshold. The coupling constant obtained by the Weinberg condition is crucial for this behavior.

\section{Conclusions}
In this work, we have studied a scalar QFT that serves as a toy model for the two-photon decays of a bound state. 
This decay is mediated by a triangle diagram, that we calculated using different approaches. In particular, we have analyzed similarities and differences between the case of a deeply bound state (pion-like), in which relativistic effects are relevant, and a loosely bound state (positronium-like), for which one `non-relativistic' pole dominates.   
In the future, one should repeat this study by using fermions as intermediate states and photons as final states. In this way, a direct connection between the positronium and the neutral pion will be possible. 

\end{document}